\documentclass[preprint,aps,11pt]{revtex4}
\usepackage{latexsym}
\usepackage{amsmath}
\usepackage{times}
\usepackage{amssymb}
\usepackage{fancyheadings}
\usepackage[T1]{fontenc}
\usepackage[utf8]{inputenc}
\usepackage{url}
\usepackage{hyperref}
\usepackage{color}
\usepackage{graphicx}
 \usepackage{epsfig}
\usepackage{mathptmx}
\usepackage{bm}
\usepackage{array}
\usepackage{algorithm}
\usepackage{algorithmic}
\usepackage{url}
\usepackage{hyperref}
\usepackage{algorithm}
\usepackage{algorithmic}
\usepackage{color}

\usepackage{lscape}
\usepackage{draftcopy}

\begin{document}

\title{Antiferromagnetic Barkhausen noise induced by weak random-field
  disorder}

\author{Bosiljka Tadi\'c$^{1,2,3}$}
\affiliation{$^1$Department  for Theoretical Physics, Jo\v{z}ef
    Stefan Institute, 1000 Ljubljana, Slovenia} 
\affiliation{$^2$Complexity Science Hub, Metternichgasse 8, 1030 Vienna, Austria}
\affiliation{$^3$Institute of Physics, Pregrevica 118, 11080 Belgrade, Serbia}

\begin{abstract}
This study numerically investigates magnetisation reversal processes driven by an external magnetic field in three-dimensional antiferromagnetic spin models with weak random field disorder. Considering an extremely weak disorder and low temperature, we observe a step-wise hysteresis loop and the appearance of short magnetisation bursts of a characteristic triangular shape; the number of bursts increases with disorder, indicative of Barkhausen-type noise. These phenomena are attributed to the simultaneous reversal at a given external field of segments composed of spins with identical neighbourhoods. A local random field orients one or more spin neighbours, resulting in small, ferromagnetic-like clusters distributed throughout the system. As disorder increases, these clusters may merge to form a labyrinthine structure within the antiferromagnetic background, facilitating brief avalanche propagation. The results demonstrate that, compared with familiar random-field ferromagnets, the observed antiferromagnetic Barkhausen noise and the related avalanche sequence have a profoundly different structure, organised into peaks associated with the transition between magnetisation plateaus.  They exhibit prominent cyclical trends and disorder-dependent multifractal fluctuations, with the singularity spectrum quantifying the degree of disorder. The activity avalanches exhibit scale invariance resembling that recently found in experiments with disordered ferr\textit{i}magnets and martensites, as well as in quantum Barkhausen noise, which are associated with active geometric regions rather than individual-spin dynamics. The observed scaling behaviour is interpreted in terms of self-organised critical dynamics.
\\[3pt]
\end{abstract}
\maketitle

\section{Introduction\label{sec:intro}}
Antiferromagnetic materials are the focus of current research for their
unique features with interpenetrating sublattices and fast spin dynamics, making promising technology applications instead of traditional disordered ferromagnets with the domain structure \cite{AFM4memory_Science2016,AFspintronic_beyond2024rev,Review_MagneticSensors2024}. Particularly, weakly disordered bulk and low-dimensional antiferromagnetic materials and layered heterostructures are intensively investigated; see recent reviews  \cite{2Dmagnets_reviewYaoNano2021,2Dmagnets_reviewScience2019}.
These systems often exhibit strong spin anisotropy and are suitably
described  by random-field Ising models, with the local random fields
emanating from  underlying weak structural defects in otherwise
regular lattices.
More complex structures, e.g. self-assembled nanomaterials \cite{SAreview_nanostrMat2018appl}, can be represented by Ising spin networks, which capture the effects of geometric frustration \cite{GeomFrustr_JPhyscSI2011Gardner,Frustrated_spinsystems2024review}  as the origin of new phenomena in these complex assemblies \cite{Isingnets_BTRP25}. 
The occurrence of hysteresis behaviour in random magnets is a
fundamental feature exploited in many technological applications. 
The magnetisation-reversal processes in antiferromagnets driven by
the external magnetic field are much less investigated, in contrast to
disordered ferromagnets having domain structure, where the collective
magnetisation fluctuations are understood as linked to the
domain-walls de-pinning and propagation \cite{RFIMreview1,BarkhausenReview,RFIMreview2024}. 
Specifically, the underlying domain-walls dynamics lead to the magnetisation bursts, known as the Barkhausen noise (BHN) signal. Groups of bursts appearing between the domain-wall depinning and new pinning position make the magnetisation avalanches, which exhibit the scale-invariance associated with the out-of-equilibrium hysteresis-loop critical point; see recent review for extensive computational techniques
developed to extract the hysteresis-loop criticality from associated BHN signals \cite{RFIMreview2024} for different systems studied with realistic driving at finite temperature and demagnetisation fields. 
The BHN signal captures the collective magnetisation fluctuations along the hysteresis loop, resulting in a nontrivial multifractal structure established in experimental \cite{ShashkovSSP2014} and theoretical
\cite{tadic2016multifractal,BHNthinSciRep2019,BHNdemagRFIM2024pre}
investigations. The BHN features and scaling of magnetisation
avalanches are thus used for noninvasive monitoring of the sample's
domain structure related to random defects and its impact on the
magnetisation processes.

In contrast to random ferromagnets, the magnetisation fluctuations
phenomena on the hysteresis loop in disordered antiferromagnetic systems are much less investigated.
It is interesting to note that  in antiferromagnets with structural disorder, the impact of these nonmagnetic site defects on the system's critical behaviour is  adequately described by the random-field
antiferromagnetic model, theoretical and experimental studies demonstrate
\cite{AF2RFIM_Cardy1984,AF_pure-rn3DIsing_Cowley_1980exp,AF2RF_KleemanPRL1995,AF2RF_DWalls1998Klee,AF2RF_dynamicsPRB1993exp,AF_RF_SciRep2020exp}. 
Meanwhile,  the addition of site defects leads to nonuniversal
critical features of the hysteresis-loop criticality in the
ferromagnetic random-field Ising model, as shown in  \cite{BHNsitedefectsPRL1996}.
Experimental studies of the antiferromagnetic material
${\mathrm{PrVO_3 }}$ with \textit{weak random-fields disorder} in \cite{AF2RF_HLinPrVO3}
revealed a characteristic structure of the hysteresis loop with
plateaus. A similar hysteresis loop structure was observed in the experiments with disordered ferimagnet ${\mathrm{DyFe_3}}$ with Ga dopped Fe sites
\cite{BHNrFiMexp_PRE2023}, in pyroxene ${\mathrm{CoGeO_3}}$ \cite{AFplateau_exp2021pyroxene}, and many other materials. Theoretically, the origin of the hysterisis-loop plateaus can be both quantum and classical 
\cite{AF2RF_HLtheoryPRB2011Shukla,Takigawa2011}; it is often connected
to \textit{geometric frustration} effects induced by complex structure and interactions in networks and lattice spin models \cite{HOC_HLentropy2020,GFrustr_graphTh_PRL2025,GeomFrustr_Ronceray2019PRE}.

The Barkhausen-type noise was experimentally observed in the antiferromagnetic-ferromanetic (AF/FM) bilayers \cite{ShashkovSSP2014} and   Ga dopped ${\mathrm{DyFe_3}}$ ferimagnet \cite{BHNrFiMexp_PRE2023}, the resistance fluctuations of the antiferromagnetic thin film
 \cite{AFresistsnceBHN_EPL2012} and in simultaneous magnetic BHN and
 acoustic emission during the thermally induced martensitic transition in Ni-Ma-Ga single
crystals \cite{BHNmartensites_BCN2013}. Recently, quantum Barkhausen-type
noise was reported in experiments with ${\mathrm{LiHo_{0.4}Y_{0.6}F_4}}$ single
crystal at low temperatures, associated with correlated quantum tunnelling phenomena \cite{QuBHN_PNAS2024}.
Experimental study in \cite{AF_DWstructureExp2024} reports  a domain-wall structure that arises in antiferromagnets ${\mathrm{MnF_2}}$ and ${\mathrm{Cr_2O_3}}$ crystals during the first-order spin-orientation transition under the influence of the external magnetic field.
Considering more complex structures as networks that model self-assembled nanomaterials
\cite{AFModelPRL2005,tadic2021hidden,Isingnets_BTRP25}, the field-driven reversal of Ising spins with antiferromagnetic coupling   along the
network's edges show BHN-type response; the associated magnetisation
avalanches exhibit universal scaling  exponents in the class of mean-field self-organised criticality  \cite{tadic2024fundamental}, resulting from the critical branching processes \cite{SOC-MFexp1988}. 
(Note that a different scaling in the class of directed sandpile automata \cite{SOC_dirSPA1989}
correspond to the higher-order interactions embedded in triangles
\cite{tadic2024fundamental}). 
In contrast to the hysteresis-loop criticality in the random-field Ising
model on compact lattices described above \cite{RFIMreview2024}, this
scaling behaviour of antiferromagnetic systems appears to be induced
by the complex network's environment \textit{without any magnetic
  disorder}, and appears as a manifestation of a self-organised
critical behaviour \cite{SOCweDynamics2021}.
Conversely, in the regular crystalline structures, the above-mentioned experimental findings in the antiferromagnetic materials suggest that the role of weak disorder could be crucial.
Given the regular sub-lattices structure and their response to the external magnetic field, entirely different mechanisms govern the magnetisation reversal in random antiferromagnetic samples, compared to those in random ferromagnets.  Hence, the genesis of the BHN-type response and the nature of criticality on the hysteresis loop in antiferromagnets with compact lattice structure and  weak disorder, as appropriate
models of these experimentally studied systems, is yet to be understood.   

In this work, we study the field-driven magnetisation reversal in a three-dimensional antiferromagnetic system of Ising spins and weak random field disorder.
Using the zero-temperature numerical simulations, we demonstrate how the magnetisation bursts appear with gradually increasing weak disorder,  leading to a specific type of Barkhausen noise and step-wise hysteresis loop. These bursts organise into uneven, well-resolved peaks associated with the transition between the magnetisation plateaus. 
Detailed analysis of the observed noise reveals that the nature of
fluctuations induced by weak random fields is different compared to
the ferromagnetic random-field Ising model, and the identified
magnetisation activity avalanches have different scaling behaviour, similar to one experimentally observed in doped ferimagnets and quantum BHN. 
The avalanche sequences in time have cyclical trends with
multifractal features and their singularity spectra, reflecting the strength of disorder, can be used to quantify the state of the underlying spin system.

\section{Model, Simulations and Methods for the Signal Analysis\label{sec:methods}}

\textit{The model of weakly disordered anti-ferromagnetic system}  is defined on the 3-dimensional regular cubic 
lattice with Ising spins $s_i=\pm 1$ associated with the lattice
sites $i=1,2,\cdots 10^6$ with the nearest-neighbour antiferromagnetic
coupling and weak random fields. The Hamiltonian is given by
\begin{equation}
{\cal{H}} = -\sum _{i,j}J_{ij}s_is_j -\sum_{i}h_is_i- H_t\sum _is_i  \ ,
\label{eq:ham}
\end{equation}
where the spin-spin coupling is $J_{ij}=-J$, indicating antiferromagnetic interactions. At each site, the spin interacts
with a quenched random fileld $h_i$,  taken from
Gaussian distribution $\rho(h)=e^{-h^2/2f^2}/\sqrt{2\pi}f$ of zero mean and the variance
$\langle h_ih_j \rangle=\delta_{i,j}f^2$.
The external field
$H_t$ coupled to all spins varies with time $t$ from large negative to
large positive values and back to complete the hysteresis loop, as
explained below.\\

\textit{Simulations of the spin-reversal dynamics} are performed by suitably
adapting  the deterministic zero-temperature  dynamics algorithm, in analogy to the ferromagnetic random-field Ising model (RFIM) \cite{RFIMreview1,BarkhausenReview,RFIMreview2024}. Specifically, a spin $s_i$  flips $s_i(t+1)=-s_i(t)$
to align with its local field $h_i^{loc}=\sum_{j}J_{ij}s_j+H_{t}+h_i $, which consists of the field
provided by the spin's six nearest neighbours $h_i^{nn}=\sum_{j}J_{ij}s_j$, time varying external
field  $H_t$ and quenched local random field $h_i$. 
The spin system is slowly driven by ramping the external field
$H_t\to H_t+r$ starting from the uniform state $\{s_i=-1\}$ along
the ascending branch of the hysteresis to complete the magnetisation
reversal, and  similarly $H_t\to H_t-r$ along the descending branch to close the loop. Here, the parameter $r$ controlling the driving rate  is kept constant.  
In analogy with random ferromagnetic systems \cite{RFIMreview1,RFIMreview2024}, we employ the \textit{adiabatic
driving}, which respects the avalanche propagation. See a detailed
program flow in Appendix \ref{alg:dynamics}.
Specifically, the external field is kept constant until the triggered activity avalanche stops, at which point the field is increased again. 
At each time step $t$, the whole system is updated in parallel,
searching for the unstable spins where $s_ih_i^{loc}<0$ and their
flips to align  with the local field are accomplished with a
probability $p$. Note that $p=1$ signifies the strictly deterministic
zero-temperature dynamics, used in the ferromagnetic RFIM \cite{RFIMreview1,BarkhausenReview,RFIMreview2024}.
However, this rule is not directly applicable in antiferromagnetic
systems with the two mutually interpenetrating sub-lattices, where the surrounding spins belong to the sub-lattice with oppositely orientated spins; thus, situations where the local field is zero, leading to spin frustration and possible back flips, may occur.  
To prevent an infinite loop where a spin flips back and forth, we set
a probability $p\lesssim 1$ for the spins to align with the local
field. A similar idea was used in modelling switching current in ferroelectrics \cite{FE_Shur1999,FEbhn_BT2002} to account for the impact of the uncompensated part of the depolarisation field. Note that, these probabilistic flips imply that the fraction $1-p$ of all spins
may not be reversed along a hysteresis branch (see sec.\ \ref{sec:hl}). 
Similar effects can also be attributed to a finite low temperature.
In this work, we have $p=0.95$ if not otherwise
specified; we also fix the driving rate $r=0.008$. 
The system's size  $V=10^6$ spins and periodic boundary conditions are
applied for better resolving groups of spins relevant to the
hysteresis steps, as explained in sec.\ \ref{sec:hl}. 
We monitor the values of the external field $H_t$, the number $n_t$ of
spin flips at time $t$, whose sequence represents \textit{the
  antiferromagnetic Barkhausen noise (AF-BHN)} signal,  and determine
the changes of total (un-normalised) magnetisation $M_t=\sum_{i=1}^{V}s_i$ with
time. 
The \textit{spin activity avalanche}, as a sequence of spin-flip events between the two following updates of the external field, is identified as a segment of the BHN  between its two consecutive intersections
$t_s$ and $t_e$ with the baseline $n_t=0$. Hence, the avalanche size
$S$ and duration $T$ are determined as $S=\sum_{t=t_s}^{t_e}n_t$ and
$T=t_e-t_s $. The structure of  the AF-BHN and accompanying
avalanches is analysed in detail in sec.\ \ref{sec:str-aval}, using the methodology described below. \\

\textit{Cyclical trendsof the AF-BHN signal and sequence of avalanches}  are determined  using  the local adaptive
detrending algorithm \cite{hu2009multifractal,tadic2023evolvingcycles}. The time series of $N$ datapoints is divided into $K=N/m-1$ overlapping segments of the length $2m+1$, where neighbouring segments overlap over $m$ points.
For each segment $k=0,1,2\cdots K$, the local trend is determined
using  the quadratic polynomial fit  $y^{(k)}(mk+\ell)$, where $\ell =0,1,2,\cdots 2m$ points. 
Specifically, for segments $0<k<K$ the trend $y_c(mk+i)$ over the overlapping points $i=0,1,2 \cdots m $  is determined by combining the
polynomial  contributions in segment $k$ and segment $k+1$ as
\begin{equation}
y_c(mk+i)= \frac{i}{m}y^{(k+1)}(mk+i) +
\frac{m-i}{m}y^{(k)}(m(k+1)+i), 
\end{equation}
such that they decrease linearly with the distance from the centre of the overlapping region. 
For the initial $m+1$ points in $k=0$ and the final $m+1$ points in $k=K$ segments, the trend coincides with the polynomial fit.  
The parameter $m$ is adjusted, depending on the considered time series.\\

\textit{Detrended multifractal analysis of multi-scale cyclic} time series is
applied, following the methodology described in 
\cite{pavlov2007multifractal,kantelhardt2002multifractal} as well as its applications to the study of various types of multiplicative cascades \cite{drozdz2025multifractal}, including the characterization of the BHN signal \cite{tadic2016multifractal,RFIMreview2024,BHN-recent-adv2025}. In this context, variations of the BHN signal induced, e.g., by the disorder, driving conditions, or sample geometry, are properly quantified by the changes in the multifractal spectra \cite{tadic2016multifractal,BHNthinSciRep2019}. 
Specifically, $2N_s$ segments of the length $n$, where $N_s=\mathrm{int}(N/n)$, are defined
by dividing the profile of the considered time sequence $Y(i) =\sum_{j=1}^i(C(j)-\langle C\rangle) $ 
 into non-overlapping segments; they are enumerated as 
$\mu=1,2,\cdots N_s$, starting from the beginning of the time series,
and $\mu =N_s+1,\cdots 2N_s$, starting from the end of the time series. 
  At each segment $\mu$, the local trend $y_\mu(i)$ is found by
  polynomial fit and  the standard deviation  $F(\mu,n)$  around  the local trend is
 determined as 
\begin{equation}
 F(\mu,n) =\left\{ \frac{1}{n}\sum_{i=1}^n[Y((\mu-1)n+i)-y_\mu(i)]^2\right \}^{1/2} ,
\label{eq-F2}
\end{equation}
 for $\mu=1,\cdots N_s$, and similarly, $F(\mu,n) =\{\frac{1}{n}\sum_{i=1}^n[Y(N-(\mu-N_s)n+i)-y_\mu(i)]^2\}^{1/2}$ for $\mu
=N_s+1,\cdots 2N_s$.
Then the $q$-th order fluctuation function $F_q(n)$ of the 
segment length $n$ averaged over all segments is  given by
\begin{equation}
F_q(n)=\left\{\frac{1}{2N_s}\sum_{\mu=1}^{2N_s} \left[F^2(\mu,n)\right]^{q/2}\right\}^{1/q} \sim n^{H(q)}  \ .
\label{eq:FqHq}
\end{equation}
Its scaling properties for a range of the segment lengths $n\in[2,N/4]$ are investigated, leading to the generalised Hurst exponent $H(q)$, as indicated by the expression (\ref{eq:FqHq}).
The parameter $q$ takes on various positive and negative values, allowing for different enhancements of the large and small fluctuations, respectively, to achieve self-similarity throughout the entire time series. Having different values $H(q)$ vs $q$ represent the multifractal spectrum; meanwhile, for a monofractal, $H(q)=const$ corresponds to the standard Hurst exponent $H(2)$. The spectrum $H(q)$ can be represented in terms of a familiar \textit{singularity spectrum} $\Psi(\alpha)$ of the local singularity H\"older exponents $\alpha$. This interpretation is related to the scaling exponent $\tau(q)$ of the multifractal box probability measure \cite{pavlov2007multifractal}. Specifically, the following relations apply \cite{kantelhardt2002multifractal}   $\tau(q)=qH(q)-1$, and $\Psi(\alpha)$ is obtained by the Legendre transform   of the scaling exponent $\tau(q)$, i.e., $\Psi(\alpha)=q\alpha -\tau(q)$, where   
$\alpha=d\tau/dq=H(q)+qdH(q)/dq$.

\section{Hysteresis plateaus and the emergence of  antiferromagnetic
  Barkhausen noise in weak random fields\label{sec:hl}}
\begin{figure}[hptb!]
\begin{tabular}{cc}
\resizebox{38pc}{!}{\includegraphics{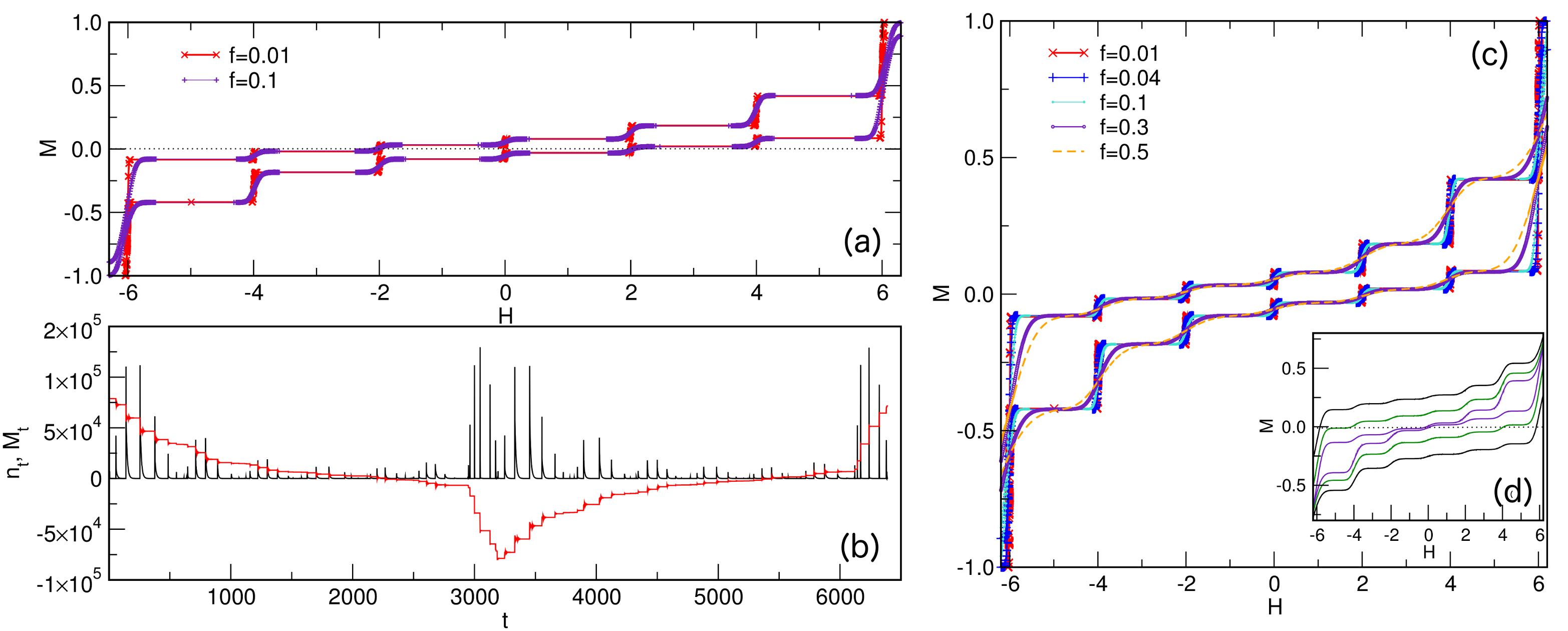}}\\
\end{tabular}
\caption{(a) The hysteresis loop $M$ vs $H$ for the antiferromagnetic
  system with a weak random-field disorder $ f=0.01$, showing the
  magnetisation plateaus and the changing transition regions between
  them for slightly increased disorder $f=0.1$.  (b) The appearance of
  magnetisation bursts $n_t$ for very low disorder $f=0.01$, and the
  magnetisation  evolution along the descending and ascending
  hysteresis branches $M_t$  (normalised to fit the scale). 
  (c) The hysteresis loops for different disorder $f$, indicated in
  the legend, and fixed probability $p=0.95$. The inset (d) shows how the hysteresis loop for fixed disorder $f=0.3$ changes when $p$ is varied from 0.995 (inner line), to 0.9 (middle line) and 0.8 (outer line).}
\label{fig:HL-f}
\end{figure}

As stated above, our objectives are to demonstrate how the antiferromagnetic Barkhausen-type noise (AF-BHN) emerges due to weak random-field disorder. Therefore, fixing the other parameters, we perform simulations at varied width  $f$ of the
random field distribution. As explained in sec.\ \ref{sec:methods}, for each value of
$f$, the spin sample is endowed with a value $h_i$ of a random field
at site $i=1,2\cdots V$, which remains quenched during the
simulation time. Meanwhile, the external field is varied
adiabatically along the hysteresis loop. At each time step $t$, the system is updated in parallel, i.e., all flipped spins are registered, and the respective changes in the local fields are stored for the next step. The external field is changed only when no flipped spins are registered. 
AF-BHN represents a time sequence of the number of spin reversals $n_t$ at time $t$, triggered by a changed value of the external field $H_t$, resulting in a corresponding change in magnetization $M_t=\sum_is_i$.  
For a very weak random field disorder, $f=0.01$, the temporal
evolution of the magnetisation $M_t$ is demonstrated along with the
number of flipped spins $n_t$ in lower left panel of Fig.\ref{fig:HL-f}; the
corresponding hysteresis loop---the magnetisation $M=\sum_is_i/V\equiv M_t/V$ vs
field $H$, is shown in the upper panel.

Fig.\ \ref{fig:HL-f} shows that, at low temperature, even very weak random fields have
considerable impact on the hysteresis shape, compared to pure
antiferromagnetic samples, where we would expect that one sublattice 
flips at $H=-6$, then holding the magnetisation $M=0$ until the field
reaches $H=+6$, where the other sublattice would flip, and vice
versa. 
With weak disorder, however,  two effects make the hysteresis loop significantly different. First, a broadening of the lines at $H_t=\mp 6$ occurs, because of small random fields at the respective sublattice sites, which shifts the counteracting external field from the exact value $\mp
6$. This effect is more pronounced by increased disorder, best
illustrated in Fig.\ \ref{fig:HL-f}d inner curve for high flipping
probability $p$. Furthermore, only a part of the respective sublattice spins fulfil the condition to reverse at $H_t=\mp 6$; the remaining spins
reverse at field values, $H_t=\mp 4, \mp2, 0$, making the
magnetisation steps and plateaus between them, as explained below, resulting in a nontrivial structure in between these two main loop parts.
Precisely, distinct groups of spins appear whose neighbourhood is differently altered by the presence of a nonzero random field. 
This effect is well pronounced at low temperature and very weak random fields, with moderate reversal probability $p$, cf.\ Fig.\
\ref{fig:HL-f}a,c. Whereas the increased disorder and higher
temperature (reduced $p$) tends to smooth these sharp hysteresis steps. More precisely, considering the ascending branch, a fraction of sublattice spins that flip at $H\sim -6$ are spins at sites that have $k=0$ flipped neighbours from the other sublattice, where the random fields are practically zero.  
The remaining spins have at least one neighbour $j$ that flipped at
$H_t=-6-h_j$, thus making the neighbours' contribution to the local
field $h_i^{nn}=4$, which requires the external field $H_t=-4$ to
fulfil the spin flip condition.  Thus, all spins in the system that have one neighbour already flipped form a group that has a chance to reverse near the field $H_t=-4$, resulting in a new magnetisation step.  
Generally, such  magnetisation steps correspond to the reversal of
segments of the system consisting of spins with the number of flipped
neighbours $k=6-j$, where $j=6,5,\cdots ,0$. They require
correspondingly larger external fields to counteract them; cf.\  Table\ \ref{tab:fmclusters} and  Fig.\ \ref{fig:HL-f}. Note that these segments contain spins at different locations in the system.

Nearby segments may be mutually connected, comprising a small
local ferromagnetic (FM)-like cluster of equally oriented spins; see Table\
\ref{tab:fmclusters} for the size of a cluster that can appear around a single spin
 with a given number of flipped neighbours. 
When the disorder is too weak, these clusters are sporadic and scattered over the lattice. However, with increasing random field
disorder, they may join together, making a complex branched structure
in the antiferromagnetic background. These clusters then 
support propagation of spin activity, in analogy to avalanches in random ferromagnets.
Furthermore, the spin flips with a limited propagation cause the short
magnetisation bursts $n_t$ that appear at the
transitions between two consecutive hysteresis plateaus. These magnetisation bursts are reminiscent of the BHN in
random-field ferromagnetic samples, as mentioned above. However, the
structure of the antiferromagnetic BHN is considerably different, as
we explain in the following section. 
\begin{table}[!]
\caption{The number of random-field assisted flipped neighbours of a spin in $(\pm x,\pm y,\pm z)$ direction and the impact
  $h_i^{nn}=\sum_jJ_{ij}S_j$ on the local field,  requiring
  counteracting external field $H_t^{ca}$; the corresponding local FM-like cluster size in 2D and  3D cubic lattice. Examples are for the hysteresis ascending branch.}
\centering
\begin{tabular}{c|ccc|ccc}
No.flipped nn&\multicolumn{3}{c}{2D lattice}&\multicolumn{3}{c}{3D lattice}\\
\hline
  &$h_i^{nn}$&$H_t^{ca}$&FM-like cluster&
$h_i^{nn}$&$H_t^{ca}$&FM-like cluster\\
\hline
0&4&-4& no cluster&6&-6& no cluster\\
1&2&-2&5&4&-4&7\\
2&0&0&9& 2&-2&13\\
3&-2&2&11&0&0&17\\
4&-4&4&13& -2&2&21\\
5&-&-&-&-4&4&23\\
6&-&-&-&-6&6&25\\
\hline
\end{tabular}
\label{tab:fmclusters}
\end{table}

 Here, we first describe the impact of the increasing disorder on
the hysteresis loop shape, as demonstrated in Fig.\ \ref{fig:HL-f}c. A larger width $f$ of the random-field distribution implies a larger number of spins
having the favouring random-field orientation to flip just before
(after)  the external field gets the corresponding $H_t^{ca}$ value for
a given hysteresis step. Consequently, the transition between two
successive plateaus is no longer sharp, but smeared over a finite
distance around $H_t^{ca}$; for example, see  Fig.\ \ref{fig:HL-f}a,
comparing $f=0.01$ with a small value $f=0.1$. For larger values,
$f=0.3$ and $f=0.5$, shown in the panel Fig.\ \ref{fig:HL-f}c, we observe that the smearing of the hysteresis steps extends over the entire plateau region, making the smaller steps difficult to distinguish. 
In the inset, Fig.\ \ref{fig:HL-f}d, we show how the hysteresis shape for
$f=0.3$ changes with the increasing fraction of unreversed spins (decreasing the parameter $p$), e.g., due to the spin frustration and finite temperature, as mentioned above. Theoretically, at zero temperature (see exact results for 1D lattice case in \cite{AF2RF_HLtheoryPRB2011Shukla}), the hysteresis loop consists of two narrow symmetrical parts that join at the point $M=0$ for $H=0$. Such a situation is illustrated by the inner line in  Fig.\ \ref{fig:HL-f}d for our model, where $p=0.995$ leaving only $0.5\%$ of unreversed spins. For decreasing $p$, however, the loop broadening occurs, increasing the distance between branches, and the shape looks more similar to ferromagnetic samples, apart from the two most significant steps at the beginning of each branch, resembling the original antiferromagnetic loop.  We note that such a situation often arises in experiments where both the finite temperature and spin frustrations can contribute.

\section{The structure of  antiferromagnetic Barkhausen noise \&
  avalanches\label{sec:str-aval}}
\subsection{Cyclical trends of the AF-BHN and the avalanche sequence}
\begin{figure}[hptb!]
\centering
\begin{tabular}{cc}
\resizebox{30pc}{!}{\includegraphics{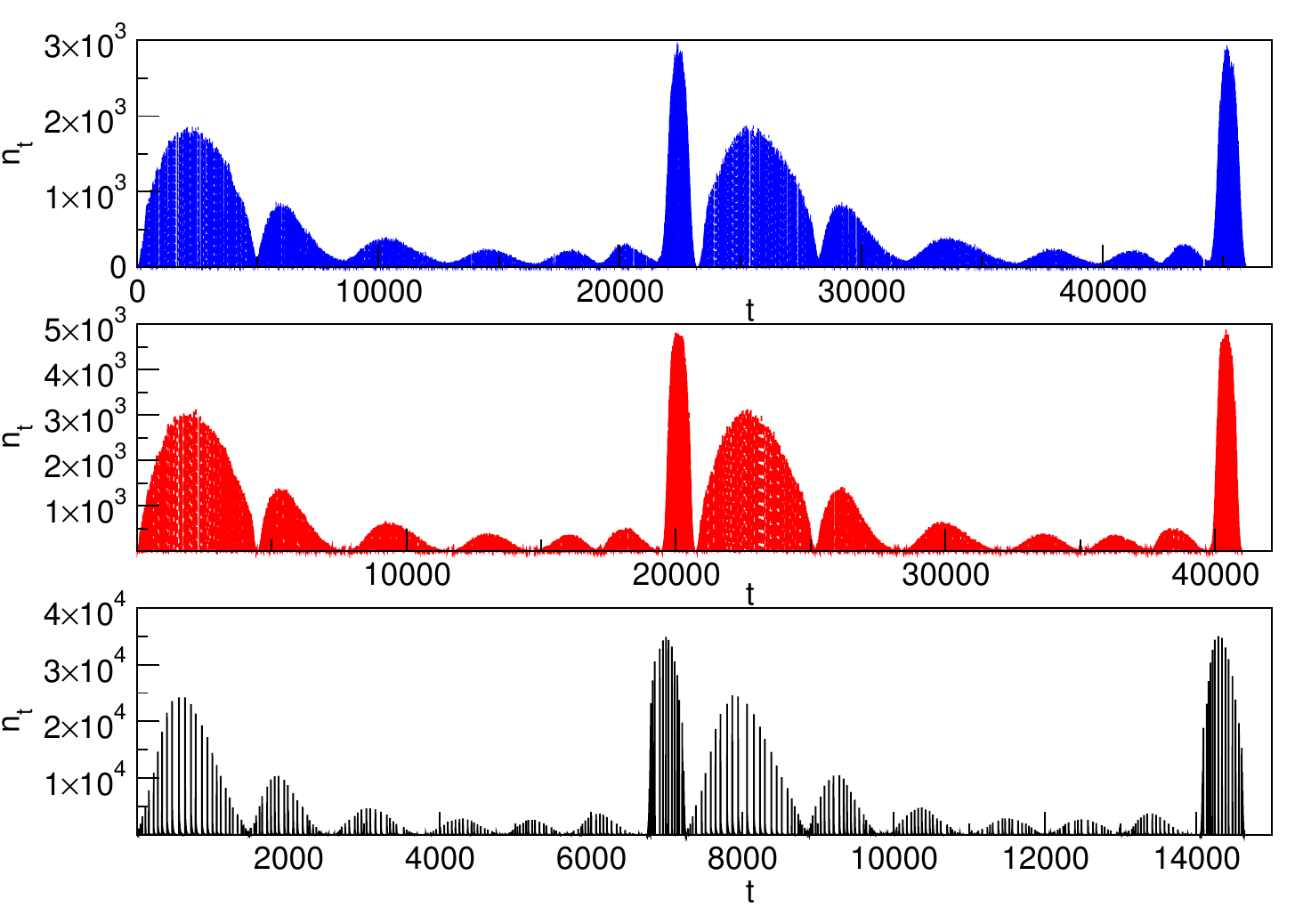}}\\
\end{tabular}
\caption{The Barkhausen noise signal $n_t$ vs time $t$ recorded during the whole
  hysteresis loop for varied  disorder $f=$ 0.04,
  0.3, and 0.5 (bottom to top panel), 
exhibiting seven peaks along each branch, but with different scales and durations. 
From left to right, seven peaks correspond to the reversal of the system's
segments comprising of spins with $k=0,1,2,\cdots 6$ reversed nearest
neighbour spins, leading to the magnetisation steps along the
ascending branch at the external
field $H_t^{ca}$ indicated in the Table\ \ref{tab:fmclusters}. Similarly, the
following seven peaks are observed at these fields in the reversed
order along the descending branch. }
\label{fig:nt3xf}
\end{figure}

\begin{figure}[hptb!]
\begin{tabular}{cc}
\resizebox{38pc}{!}{\includegraphics{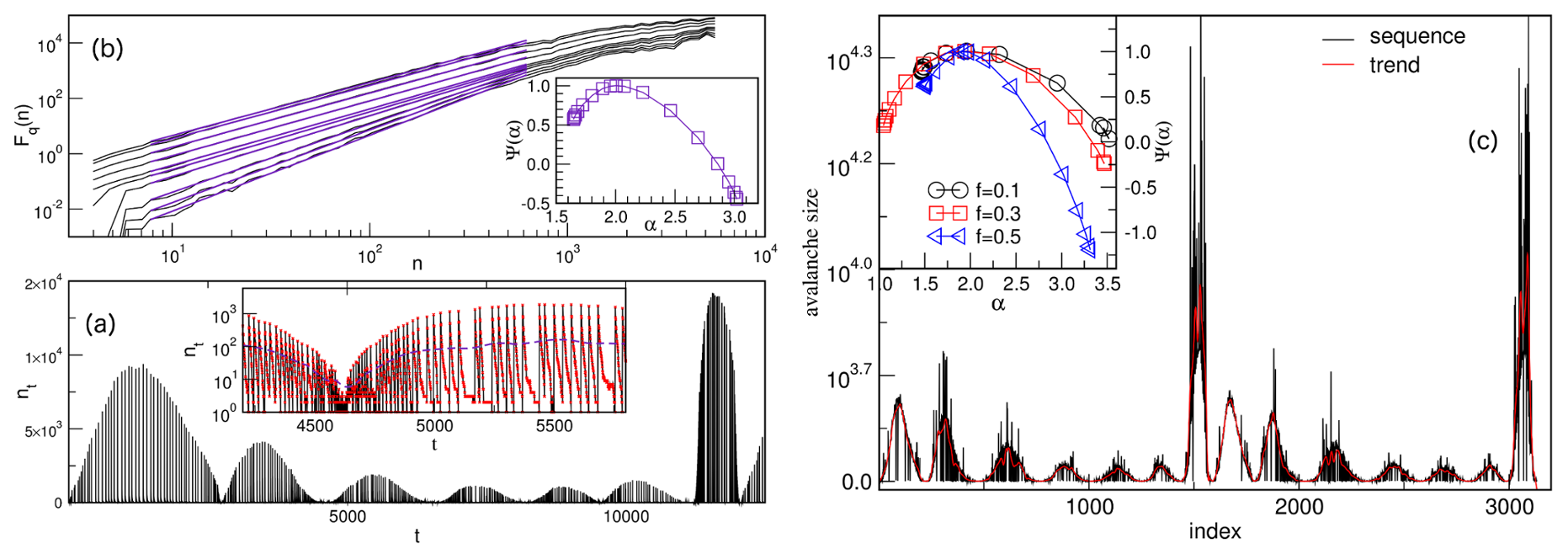}}\\
\end{tabular}
\caption{(a) The main panel shows the AF-BHN signal for $f=0.1$
  exhibiting characteristic groups of different sizes registered at
  the descending branch of the hysteresis loop; inset shows a close-up
  view of the signal between the second and third peak and its trend,
  indicated by dashed line. (b) The fluctuation function $F_q(n)$ vs. time interval $n$ of the whole loop signal's trend; each line
  corresponds to a different $q\in[-4.5,4.5]$, every second line is
  shown for better vision; the thick straight lines indicate the scaling region where the generalised Hurst exponent is determined. The inset shows the  corresponding singularity spectrum $\Psi(\alpha )$ vs
  $\alpha$. 
(c) The main panel shows the sequence of avalanche sizes for $f=0.3$ (black line)  and its cyclical
  trend (red line). Inset: the singularity spectrum $\Psi(\alpha)$ vs
  $\alpha$ of the avalanche-sequence trend for three values of the
  random-field disorder indicated in the legend.}
\label{fig:trend_nt}
\end{figure}

As explained above, the presence of quenched random fields leads to the
hysteresis steps and the appearance of local FM-like clusters,
through which the spin reversal activity can propagate, resulting in
short magnetisation bursts. The temporal sequence of these bursts
makes the antiferromagnetic BHN, which we examine in the following;
cf.\ Figs.\ \ref{fig:nt3xf}-\ref{fig:trend_nt}.

Remarkably, these magnetisation fluctuations are organised in seven
peaks, which are well-resolved at low temperatures and weak disorder,
along each hysteresis branch, as shown in Fig.\ \ref{fig:nt3xf};
each peak associates with a distinct group of spins with respectively $k=0,1,2,\cdots 6$ reversed nearest
neighbours, 
corresponding to a given step in the hysteresis in Fig.\
\ref{fig:HL-f} at the external field
$H_t=H_t^{ca}$ indicated in Table\ \ref{tab:fmclusters}. 
Precisely, for a very weak disorder, the small isolated clusters are 
at different locations in the lattice. Then, at the $H_t=H_t^{ca}$ corresponding to a given step on each hysteresis branch, spins in these clusters can flip simultaneously, resembling an additional sublattice, which leads to a large jump in the signal. The activity may briefly propagate in a local cluster. This situation leads to
a sequence of high and short bursts,  for example, those visible in Fig.\
\ref{fig:HL-f}b. 
With increasing disorder,  a larger number of bursts and potential propagation over a labyrinth of connected local clusters may occur, leading to a longer and more complex signal; cf.\ the inset to Fig.\ \ref{fig:trend_nt}a. Consequently, the magnetisation reversal process takes longer, and the individual bursts are lower in height. Meanwhile, a gradual broadening of each of the seven peaks is evident when comparing different cases in Fig.\ \ref{fig:nt3xf}. At the same time, the overlap between successive peaks increases, in accordance with the smearing of the transition between hysteresis steps, which is demonstrated above in
Fig.\ \ref{fig:HL-f}c.  These changes in the global structure of the
magnetisation noise can be quantified by studying modulations of its cyclical trend, as demonstrated below; see Fig.\ \ref{fig:trend_nt}.

Using the methodology described in sec.\ \ref{sec:methods}, we first determine the cyclical trend of the signal along the whole hysteresis loop and then study its multifractal features. The example in Fig.\ \ref{fig:trend_nt}a is for the case $f=0.1$; a part of the signal is shown for better vision.  
The trend closely follows the peak-like structure of the AF-BHN
signal, resulting in an irregular cycle, whose modulations are 
quantified  by the multifractal spectrum. Precisely, the generalised fluctuation
function, $F_q(n)$ vs time interval $n$,  defined in (\ref{eq:FqHq}),
is computed for the signal's trend for different values of the scale
parameter $q$. Its plot in Fig.\ \ref{fig:trend_nt}b
shows multifractal features in an extended range of time
intervals, where the lines for varied $q$ have different slopes
defining the spectrum of the generalised Hurst exponents $H(q)$. The
$H(q)$ vs $q$ dependence is then transformed (see sec.\ \ref{sec:methods}) into a singularity spectrum $\Psi(\alpha)$ vs $\alpha$, where different  $\alpha$ values suggest altered scaling for sets of datapoints along the trend. 
The spectrum is shown in the inset to Fig\ \ref{fig:trend_nt}b; it appears to be broad and asymmetrical, with an extended right side corresponding to small-scale fluctuations. 
The occurrence of modulated cycles in the AF-BHN is also manifested in
the sequence of avalanches that the signal defines. We recall that an
avalanche comprises the area covered by individual bursts between two
consecutive drops of the signal to the baseline; the close-up of the signal in the inset of Fig.\  \ref{fig:trend_nt}a.   Fig.\ \ref{fig:trend_nt}c shows an
example of the avalanche sequence, derived from the signal for
$f=0.3$, and its irregular cyclical trend. The multifractal analysis of cyclical trends of the avalanche sequences is done for  different disorder strengths. The results of their singularity 
spectra are shown in the inset to Fig.\ \ref{fig:trend_nt}c for
varied distribution $f$ of the random fields, listed in the
legend. The observed variations in these singularity spectra, therefore
serve as a quantitative indicator of the influence of disorder
strength on the hysteresis loop dynamics.
These spectra are also asymmetrical, in analogy to the ones of the signal's trends. Moreover, the spectra are gradually shrinking with
the increasing disorder, which suggests a reduced difference between
the scales of avalanches. These findings will be more precisely
demonstrated in the following section.

\subsection{Scale-invariance of antiferromagentic avalanches at weak
  disorder\label{sec:avalanches}}
\begin{figure}[hptb!]
\begin{tabular}{cc}
\resizebox{38pc}{!}{\includegraphics{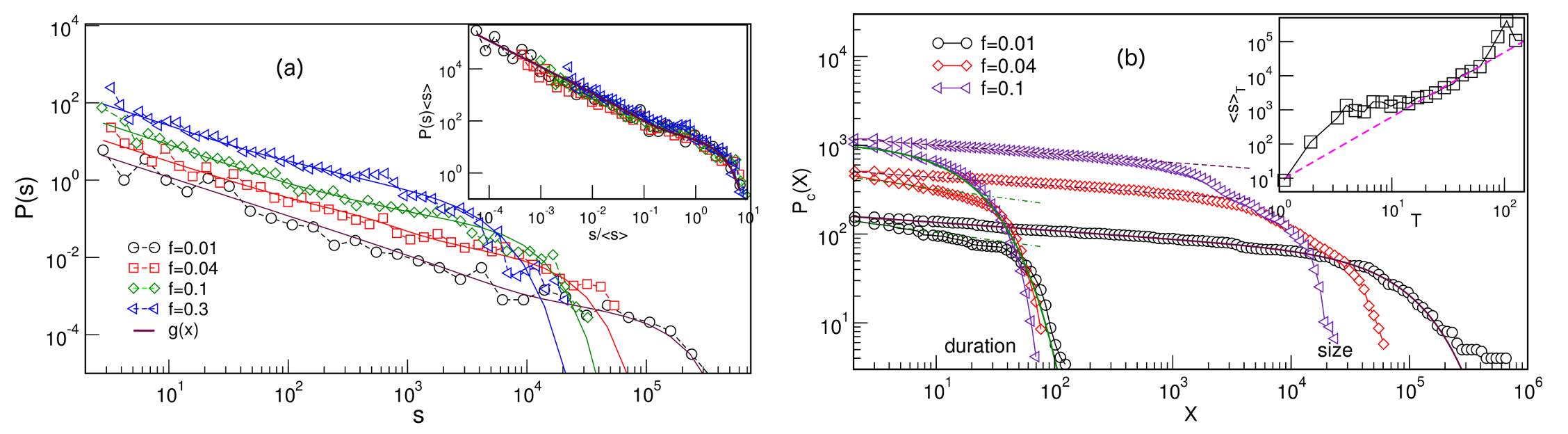}}\\
\end{tabular}
\caption{(a) Main panel shows the differential distributions $P(s)$
  of the avalanche sizes $S$ obtained from magnetisation bursts along
  the entire hysteresis loop for different disorder strength $f$,
  indicated in the legend; each line is fitted by the function
  (\ref{eq:gx}) with appropriate parameters. Inset: the scaling
  collapse of these distributions according to the expression
  (\ref{eq:scaling}), with the corresponding average value $<S>$ of the avalanche size.  
Fit line  by the expression (\ref{eq:gx})
with parameters $a_1=12,a_2=10, h=0.2,d=0.9, c=1.8,
\kappa=0.65,\sigma=1.3$. (b) Sample-averaged cumulative
distribution of the avalanche size (longer lines) and duration
(shorter lines with the same symbol) collected along the ascending
hysteresis branch for varied disorder  $f$, indicated in the
  legend. The whole curve fit, shown by thick full line  for
  $P(S)$ at weak disorder, gives the exponent $\tau_s-1=0.09$ and
  approximately the same slopes indicated by
  dashed lines apply for the straight segments of the curves
  at larger disorder, but their cut-offs are no longer
  exponential. Similarly, for the duration distribution we have
  $\tau_T-1=0.15$. Inset: average avalanche size of a given duration
  $<S>_T$ vs duration $T$ for all recorded avalanches; the scaling with the exponent $\gamma_{sT}\approx 1.9$
applies only in the intermediate durations. }
\label{fig:ddistr}
\end{figure}

The characteristic structure of magnetisation bursts with a sharp rise and gradual decay, organised into groups with cyclical trends, also manifests in the structure and statistics of the avalanches determined from the AF-BHN signal. 
For an increased (weak) disorder, triggered spin activity may
propagate beyond its local cluster through a maze of connected
clusters, leading to a large avalanche with multiple jumps and decays; cf.\ inset to Fig.\ \ref{fig:trend_nt}a. 
 These avalanches contribute to the specific form of the cut-off in the avalanche size distribution, as shown in Fig.\ \ref{fig:ddistr}.
Moreover, the restricted number of magnetisation bursts in each peak
(for a given disorder) and the peaks' mutual overlaps imply
strong restrictions on the distributions of the avalanche sizes and
durations. 
In particular, the differential distribution (normalised by the number
of avalanches) of avalanche size $P(S)$ determined from the AF-BHN signal collected along the entire hysteresis loop is shown in Fig.\ \ref{fig:ddistr}a for varied disorder strengths. These distributions exhibit a break point that separates two regions with different slopes, and a cut-off with a shoulder preceding a stretched exponential decay.

Note that this form of the avalanche-size distribution cut-off was derived in the renormalisation group theory \cite{RG_avalnchedistr2009PRE,RG_aval2024hyperuniformity}  for depinning of an elastic line or interface in a random-field landscape. 
In the present study, we have the two-slope avalanche size $S$ distributions that are fitted
by the function
\begin{equation}
g(S)=(1-\tanh(S/h))a_1S^{-1} +
\tanh(S/h)a_2S^{-0.8}\exp{[(S/d)^\kappa-(S/c)^\sigma]} \ .
\label{eq:gx}
\end{equation}
A similar expression is known for the distributions of avalanches propagating in thin
samples of the ferromagnetic RFIM \cite{BHNthinSciRep2019}, where the
first and the second slope, in general, $\tau_1$ and $\tau_2$,  take
the values known for the avalanches  of the 3D and 2D case, respectively. 
In \cite{BHNthinSciRep2019}, two slopes and these cut-offs are explained by the motion of domain walls in the restricted space of thin samples and the enhanced role of the open boundary.
 In the present case, we deal with the compartmental separation of
 the sample into FM-like clusters within an antiferromagnetic
 environment, leading to different exponents, in particular,  $\tau_1=1$ and $\tau_2<1$ in eq.(\ref{eq:gx}), which are confirmed (within numerical error bars) by respective fits. 

Another marked feature of these distributions is that they obey scale
invariance as disorder is changed, as shown in the inset of Fig.\
\ref{fig:ddistr}a. In contrast to the critical scaling of avalanches known in ferromagnetic RFIM
\cite{RFIMreview2024}, we have  entirely different scaling form
\begin{equation}
P(S,f)=\frac{R}{<S>}{\cal{P}}(\frac{S}{<S>})  \ ,
\label{eq:scaling}
\end{equation}
where $<S>$ is the average avalanche size for a given disorder $f$,
and the numerical factor $R\eqsim f/f_0$ is related to
the relative shift of the distributions along the vertical axis
(roughly, the number of detected smallest avalanches) compared to the
weakest considered disorder $f_0$. This type of scaling  is more often found in
social dynamics \cite{OSNdyn_fssradicci,OSNdyn_we13}, where it is
directly related to the group-activity levels rather than any finite
size scale, and to the fact that the scaling exponent is close to one.  See more in the Discussion section.

The sample averaged cumulative distributions $P_c(X>S)\sim
S^{1-\tau_s}{\cal{C}}(S)$ of the avalanche size and the corresponding distribution of duration are shown in the right panel of
Fig.\ \ref{fig:ddistr}b.  
Note that, because of the cut-off's shoulder, no
smaller exponent  $\tau_2<1$ can be detected in this type of statistics. 
Furthermore, for very low disorder and \textit{excluding excessive avalanches},
one can fit the distribution by the familiar power-law and the exponential cut-off, which gives  the numerical values of the ($\tau_1$) exponent, in particular $\tau_s=1.09\pm 0.06$, for the size, and
 $\tau_T=1.15\pm 0.06$ for the duration distribution.  
In this limit, predominant avalanche shapes consist of a single jump and gradual decay. Interestingly, these types of avalanches
 are recently reported in  experiments \cite{SOC_tau1_martensites2025}, analysing the avalanches in the acoustic emission signal of martensites
 exposed to quasistatic elastic deformation, also leading to
 the exponent $\tau_s=1$. (Note that a different behaviour of avalanches was observed in temperature-driven martensitic transformation, see the recent experimental study in \cite{martensites_avalanchesEV2017}). 
A similar exponent $\tau_s=1$ was reported in SOC automata models
\cite{SOC_tau1_2022logN} and  the avalanches associated with SOC on the hysteresis loop of the infinite-range spin-glass model \cite{SOC_tau1_SGlri1999}.
We also note that the exponent does not change with the increased
disorder, in contrast to the cut-offs that change dramatically, in
agreement with the above studied differential distributions. 
Moreover, the shapes of avalanches deviate from the expected scaling law
$<S>_T\sim T^{\gamma_{ST}}$, both for small and excessive sizes, as
demonstrated in the inset of Fig.\ \ref{fig:ddistr}b.  In the intermediate range, the  exponent is found close to the mean-field value $\gamma_{ST}=2$.

\section{Summary and Discussion\label{sec:discussion} }
We have numerically investigated the stochastic processes of field-driven magnetisation reversal of antiferromagnetically interacting Ising spins in the presence of weak random fields on a 3-dimensional cubic lattice with periodic boundary conditions. The weak quenched random-field disorder is varied by changing the width  $f\in [0.01,0.5]$ of its Gaussian distribution with zero mean. Our results reveal new emergent phenomena with the Barkhausen-type noise, rendering the hysteretic behaviour fundamentally different from that of a pure antiferromagnetic system, even at the weakest disorder. At the same time, these phenomena differ significantly from the well-known hysteresis-loop criticality and the Barkhausen noise in random-field ferromagnets. Specifically, we find:

\begin{itemize}
\item \textit{The Hysteresis loop exhibits six magnetisation plateaus} along each hysteresis branch between the external field values $H=-6,-4,-2,0,2,4,6$. They are related to the occurrence of sets of spins with $k=1,2,3,4,5$ random-field-assisted flipped neighbours, thus reducing their contributions to the local field such that it can be counteracted by these external field values along the ascending branch, respectively.
  At vanishing disorder, $k=0$ defect neighbours require $H=-6$ to flip one sublattice, which makes $k=6$ flipped neighbours and  $H=+6$ for the spins in the other sublattice. Moreover, the set of spins with a similar neighbourhood is not a connected area; instead, they are scattered at different locations in the system.
By increasing disorder, the transition between consecutive plateaus is
 smeared. For vanishing temperature and geometric frustration (the spin alignment probability $p\to 1$), the reversal is complete, and the loop shape is ideally thin in the centre $(M=0, H=0)$. However, with a finite percentage $1-p$ of unreversed spins, the loop is broadening and receives nonzero magnetisation at $H=0$, as often seen in the
experiments with weakly disordered antiferromagnets.

\item \textit{The Antiferromagnetic Barkhausen-type noise emerges} with the
  magnetisation bursts that are associated with the
  transitions between the consecutive hysteresis-loop  plateaus; hence, they are organised in seven well-resolved peaks containing groups of the magnetisation bursts of different sizes. By increasing
  disorder from extremely weak $f=0.01$ to relatively strong $f=0.5$ in these systems, the structure of the AF-  BHN signal changes considerably. In particular, the peaks overlap and the number of elementary bursts per peak increases, while their heights decrease, and the overall reversal time is prolonged. 

\item \textit{The irregular cyclical trends appear} as a prominent
  characteristic of the temporal evolution of the  AF-BHN signal and the
  associated sequence of avalanches. Modulation of these cycles by the
  appearance of higher modes corroborates the increased disorder
  strength; the multi-fractal singularity spectra appropriately
  quantify these modulations. Thus, the presented multi-fractal
  methodology of the AF-BHN signal processing can be used to identify
  the level of disorder that impacts the hysteresis dynamics in weakly
  disordered antiferromagnetic systems.

\item  \textit{The magnetisation avalanches indicate SOC behaviour} without any critical disorder point. The avalanches determined from the AF-BHN exhibit the scale-invariant distributions with two slopes, where the dominant slope has the scaling exponent close to $\tau_s=1$ independent of disorder; characteristic cut-offs with a shoulder preceding the stretched-exponential decay systematically change with the disorder, and satisfy a different type of scaling collapse, which is formally similar to the one associated with the group-activity dynamics in social systems.
\end{itemize}

Mechanisms underlying these phenomena can be  linked to
the appearance of local FM-like clusters, containing equally oriented
spins due to the impact of a local random field enabling the spin flip at the boundary of two sublattices. The cluster size, see table\ \ref{tab:fmclusters}, can vary depending on the number $ k>0$ of such neighbours, where $k=0$ corresponds to the case of an ideal (pure antiferromagnetic) two-sublattices structure. The clusters represent additional compartments occurring at different locations in the system; the clusters with a given $k$ reverse simultaneously when the external field reaches a value sufficient to counteract the local field due to spin neighbours slightly shifted up or down by local random fields.
By increasing disorder, these local FM-like clusters may merge into a complex maze embedded in the antiferromagnetic  environment, enabling the propagation of spin activity and leading to larger avalanches. 
These collective behaviours, mutually separated by a stable  surrounding, lead to the above-mentioned specific shape of magnetisation bursts in the AF-BHN signal and avalanches derived from it. Regarding their geometry and statistics, these avalanches significantly differ from the critical avalanches in the random-field ferromagnetic samples; see a comprehensive review of different cases in \cite{RFIMreview2024}.   At the global level, these mechanisms lead to a characteristic shape of the AF-BHN signal and sequence of avalanches endowed with modulated cyclical trends.
On the other hand, we note that the scale invariance with characteristic scaling exponent $\tau_s=1$ of the avalanches in weakly disordered antiferromagnetic systems shows  remarkable similarity with the ones recently observed in experiments in disordered ferimagnetic materials \cite{BHNrFiMexp_PRE2023} where antiferromagnetic coupled clusters of spins play a decisive role, and in quantum magnetic systems  \cite{QuBHN_PNAS2024}, where quantum BHN is attributed to co-tunnelling of segments of domain walls. Furthermore, the shapes of the AF-BHN bursts are similar to those recently studied in inertia-driven martensites, where a similar value of the scaling exponent is also found \cite{BHNmartensites_BCN2013}. In these materials, avalanche criticality is studied at thermal-driven phase transitions; see recent work \cite{martensites_avalanchesEV2017} and references there. In these materials, it is known that segments of the high-temperature phase order together due to fast quenching.
Concerning the hysteresis-loop criticality  in weakly disordered antiferromagnets, the scaling behaviour of  the avalanche distributions for varied disorder indicates a different type of scaling function and the absence of any critical disorder, suggesting the SOC-type criticality, in analogy to the infinite-range spin-glass model \cite{SOC_tau1_SGlri1999} and Ising-spin networks \cite{Isingnets_BTRP25}. 
It is also worth noting that deterministic abelian sandpile automata driven by the
open boundary, which serves as a paradigm of SOC, have the avalanche
size exponent $\tau_s=1$ after the finite-size effects are properly taken into account \cite{SOC_tau1_2022logN}.

Note that, in this work, we considered fully periodic boundary
conditions to emphasise the role of internal disorder-induced structure on the hysteresis shape and the emergent AF-BHN.  Meanwhile, inside the sample, an open boundary separates the FM-like
clusters from the antiferromagnetic bulk. 
The presented analysis opens several questions that remain for future study. Among these, it is certainly intriguing to investigate the fractal
structure of the mazes of FM-like clusters for different disorder strengths and how it affects the universality of avalanches both in 2- and 3-dimensional  lattices. 
Exploring the hysteresis loop SOC behaviour in regular (lattice) structures with a weak random field, in comparison with that observed in complex network structures where no magnetic disorder is needed \cite{Isingnets_BTRP25}, could further reveal the specific role of geometry in the hysteresis phenomena of antiferromagnetic systems. Furthermore, the impact of changing driving mode remais to be investigated, specifically  in view of its potential influence on the statistics of avalanches \cite{ELavalanches_drive2025protocols} and crossover from the critical point behaviour to self-organised criticality \cite{Reche_crit2SOC2008}.
More simulations of different sample shapes and open boundaries, in
analogy to random ferromagnets \cite{BHNthinSciRep2019}, and a finite
temperature would more closely mimic the experimental conditions and
enable comparisons with experimental results. We have demonstrated the
hysteresis loop broadening with the impact of the low temperature and
spin frustration (here captured by a parameter $p <1$); simulations
taking thermal fluctuations, in conjunction with increasing driving
rate, as done in ferromagnetic RFIM and FM-AF bilayers
\cite{BHNdemagRFIM2024pre, FM-AFbilatersPRB2025}, are needed to
quantify the impact of finite temperature and driving rates, and differentiate thermal fluctuation effects from geometric frustration in these systems.

In conclusion, technological demands for weakly disordered antiferromagnetic materials motivate theoretical investigations aimed at a deeper understanding of hysteresis phenomena in these systems. Our study of field-driven magnetisation reversal processes in antiferromagnetic coupled Ising spin systems on a three-dimensional lattice revealed unique features arising from weak quenched random-field disorder. Notably, a hysteresis loop exhibits a sequence of plateaus, and the antiferromagnetic Barkhausen-like noise appears, whose structure significantly differs from the well-known BHN in disordered ferromagnets. The AF-BHN and the accompanying sequence of magnetisation avalanches are characterised by modulated cyclical trends and scale invariance indicating SOC dynamics, in striking contrast to the hysteresis-loop criticality in  disordered ferromagnets.
These hysteresis phenomena in weakly disordered antiferromagnetic systems can be attributed to the formation of random-field-assisted clusters of spins that are equally oriented, which reverse simultaneously at distant locations in the system. Hence, apart from the technological aspect, the observed hysteresis phenomena of weakly disordered antiferromagnets are of theoretical importance, as they point the way to elucidate universal collective dynamics, which is dominated by active geometric regions rather than individual spins. Such dynamical phenomena characterise the hysteresis behaviour of a wide class of systems, including disordered ferrimagnets and martensites, and quantum Barkhausen noise.

\acknowledgments
This work was supported by the Slovenian
Research and Innovation Agency under the program P1-0044.

\end{document}